\documentclass[twocolumn,showpacs,showkeys,superscriptaddress]{revtex4}
\usepackage{ifpdf}
\usepackage{graphicx}
\usepackage{amsmath}
\usepackage{amssymb}
\usepackage{slashbox}

\begin{document}
\title{Analysis of one-dimensional Yut-Nori game: winning strategy and avalanche-size distribution}
\author{Hye Jin Park}
\affiliation{BK21 Physics Research Division and Department of Physics, Sungkyunkwan University, Suwon 440-746, Korea}
\author{Hasung Sim}
\affiliation{IBS Research Center for Functional Interfaces and Correlated Electron Systems, Seoul National University, Seoul 151-747, Korea}
\author{Hang-Hyun Jo}
\affiliation{Department of Biomedical Engineering and Computational Science,
Aalto University School of Science, P.O. Box 12200, Finland}
\author{Beom Jun Kim}
\email[Corresponding author: ]{beomjun@skku.edu}
\affiliation{BK21 Physics Research Division and Department of Physics, Sungkyunkwan University, Suwon 440-746, Korea}

\begin{abstract}
In the Korean traditional board game Yut-Nori, teams compete by moving their
pieces on the two-dimensional game board, and the team whose all pieces
complete a round trip on the board wins.  In every round, teams throw four
wooden sticks of the shape of half-cut cylinder and the number of sticks that
show belly sides, i.e., the flat sides, determines the number of steps the
team's piece can advance on the board.  It is possible to pile up one team's
pieces if their sites are identical so that pieces as a group can move together
afterwards (piling).  If a piece of the opponent team is at the new site of one
team's piece, the piece is caught and removed from the board, and the team is
given one more chance to throw sticks and proceed (catching).  For a
simplicity, we simulate this game on one-dimensional board with the same number
of sites as the original game, and show that catching is more advantageous
strategy than piling to win.  We also study the
avalanche-size distribution in thermodynamic limit to find that it follows an
exponential form.

\end{abstract}



\maketitle

\section{Introduction}
\label{sec:intro}

\begin{figure}
\includegraphics[width=0.47\textwidth]{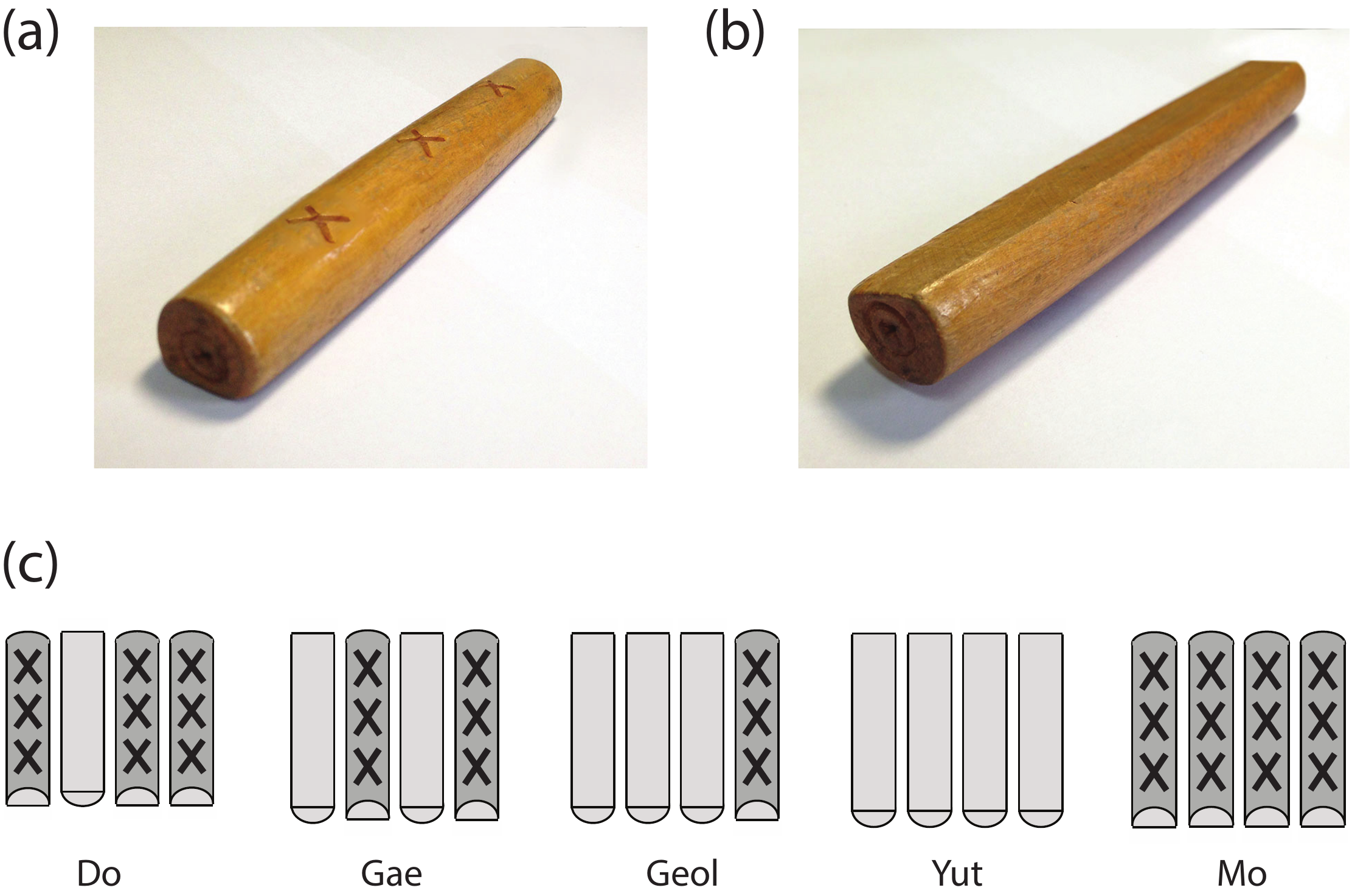}
\caption{(Color online) A Yut stick can show (a) the round back side (marked by
crosses) up or (b) the flat belly side up. A Yut stick is shaped in such
a way that the belly side has higher chance to occur than the
back side, i.e., $P_\boxtimes < P_\square \equiv p$.
(c) Four Yut sticks can have five different states, depending
on the number of belly sides: Do, Gae, Geol, and Yut for one to four belly
sides, respectively. Mo is when all four Yut-sticks are of the back sides. }
\label{fig:yut}
\end{figure}

\begin{figure}
\includegraphics[width=0.48\textwidth]{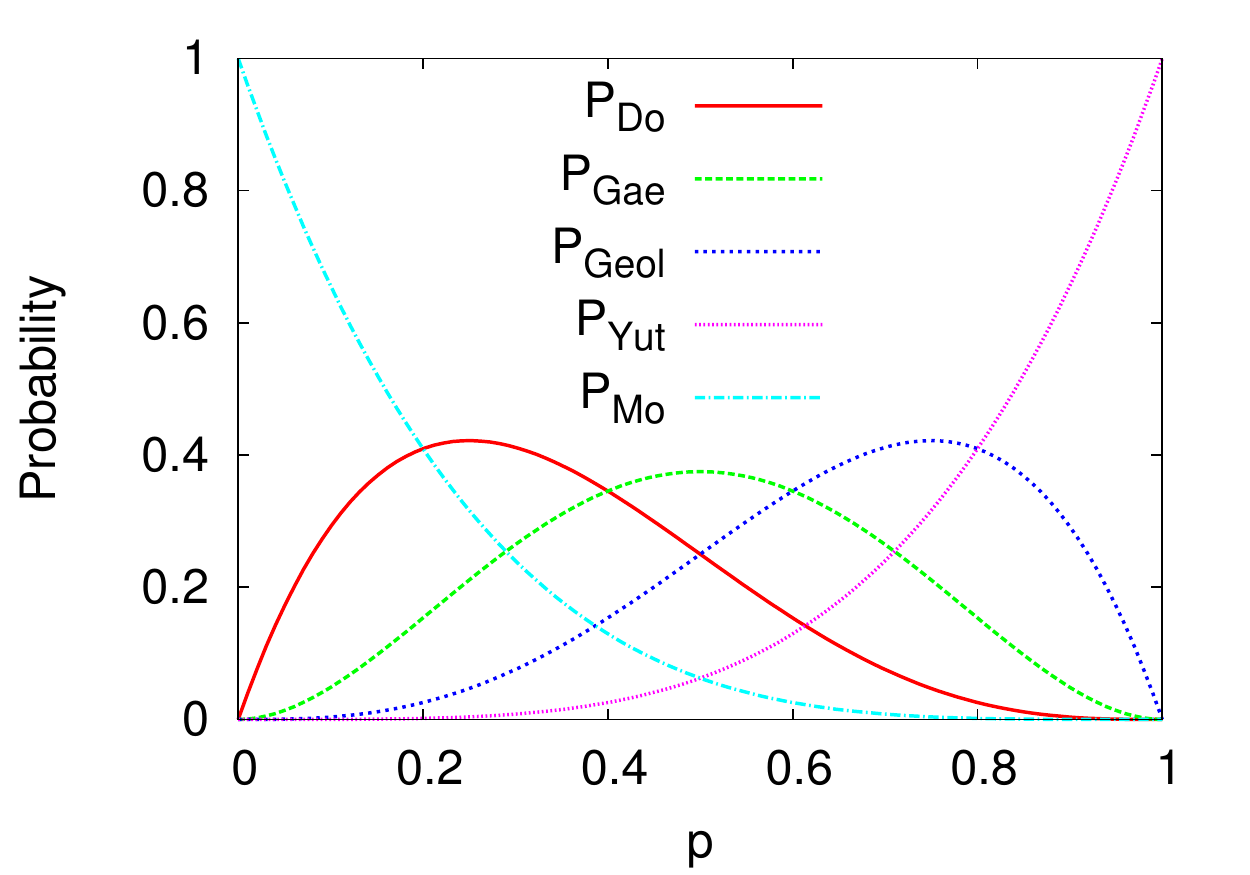}
\caption{(Color online) Probabilities for Do, Gae, Geol, Yut, and Mo in Yut-Nori
versus the probability $p$ for the belly side of one Yut stick.
Mo is the most favorable in Yut-Nori since you can proceed five steps
and you are given one more chance to throw sticks. Yut is the second best,
giving you four steps on the board and one more chance to throw.
Traditionally, $p$ is set to close to 0.6, which is easily understood
since $P_{\rm Do} > P_{\rm Yut}$ when $p < 1/(1+4^{-1/3}) \approx 0.6135$.
}
\label{fig:Pn}
\end{figure}

Yut-Nori with  Nori meaning a game, is a Korean traditional board
game.  Each team is given four pieces, usually small stones, called Mal,
and teams share four sticks called Yut and the game board called
Yut-Pan to play.  The set of four sticks acts like a dice (see
Fig.~\ref{fig:yut}): Each wooden stick has the shape of a cylinder
half-cut in a longitudinal direction and has the back side (the round side)
and the belly side (the flat side).  The number of belly sides of four
thrown sticks determines the number of steps that a piece on the board can
move.  If only one stick shows the belly side and others back sides, we call
this Do.  When one gets Do, the team can advance its piece by one step
on the game board.  Likewise, Gae and Geol are for two and three
sticks of belly sides, and the piece proceeds by two and three steps,
respectively.  If all four sticks show belly sides, which is called Yut
(the same term Yut is used both for the outcome of stick throwing
and for the name of sticks),
the team's piece is advanced by four steps, whereas for four back sided
sticks (no belly sided stick) called Mo, the piece moves by five steps
on the game board.
Yut sticks are cut in such a way that the probability $p$ for the belly side
is larger than the probability $1-p$ for the back side. For a given value
of $p$, it is straightforward to compute the probabilities for Do, Gae, Geol,
Yut, and Mo as $P_{\rm Do} = \binom{4}{1}p(1-p)^3$,
$P_{\rm Gae} =\binom{4}{2}p^2(1-p)^2$,
$P_{\rm Geol} = \binom{4}{3}p^3(1-p)^1$,
$P_{\rm Yut} = \binom{4}{4}p^4(1-p)^0$, and
$P_{\rm Mo} = \binom{4}{0}p^0(1-p)^4$, respectively.
In Fig.~\ref{fig:Pn}, we display the above probabilities as functions of $p$.
It is clear that $p$ must be larger than 0.5 to have $P_{\rm Mo} < P_{\rm Yut}$
for example. Also, if $p > 1/(1+4^{-1/3})$, $P_{\rm Yut}$ becomes larger than
$P_{\rm Do}$, which needs to be avoided since Yut is much better than Do
and thus should occur less frequently than Do. Consequently, we expect
that the probability $p$ for the belly side must satisfy $p \lesssim 0.6$.

There still remains a debate about the historical origin of Yut-Nori, but
the argument by Chae-Ho Shin, who was a Korean historian and independence movement
activist during Japan's colonization period of Korea,
has been widely accepted~\cite{wikiko}.
According to him, Yut-Nori originated from an ancient Korean kingdom Gojoseon,
which existed from 2333 BC according to the history book Samguk Yusa written
by a Buddhist monk Ilyeon in the 13th century AD. In Gojoseon era, there were five
influential tribes ruling five regions in the kingdom, and they joined
forces during wartime. The game board Yut-Pan mimics how these five tribes
spatially arranged their positions during battles. Except for the king's tribe,
all four tribes had their names from livestock; pigs, dogs, cows, and horses.
These names are reflected as the terms in Yut-Nori as Do,  Gae,  Yut, and Mo,
respectively. Chae-Ho Shin argues that the king's tribe is represented as
Geol in Yut-Nori.
It is interesting to note that the distance a piece can
move appears to correspond to the sizes and the step widths of animals.  The
speed of a pig is the smallest and a horse the fastest, giving the
increasing step size in Yut-Nori.  Even in modern Korean language, a dog is
still called Gae, and a pig Doe-Ji, probably sharing the same origin
as Do in Yut-Nori.

\begin{figure}
\includegraphics[width=0.4\textwidth]{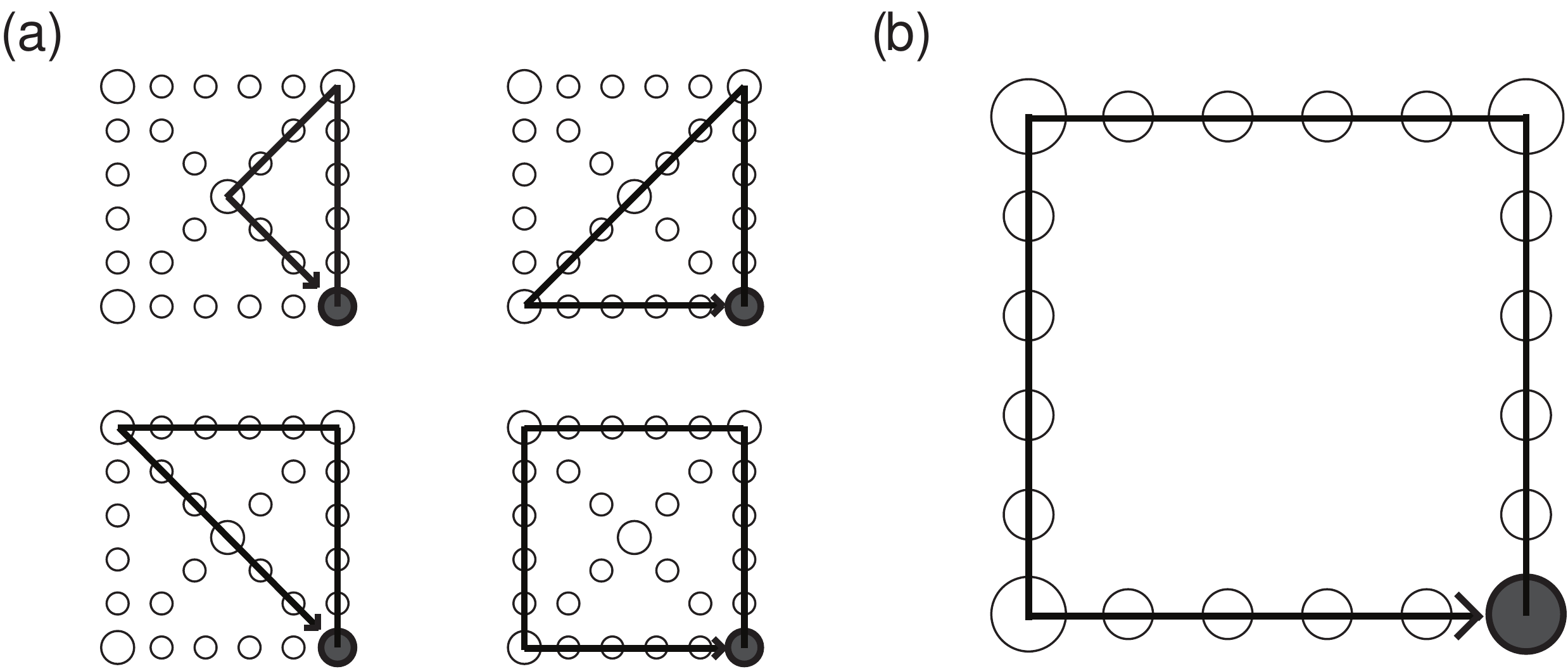}
\caption{(a) The original Yut-Pan, the game board for Yut-Nori.
Every piece (Mal) should start from the starting site marked as filled circles,
and proceeds to make a round trip around the board. When the piece is on the forks
(the top left and the top right corners, and the central position), the piece takes the shorter route.
Otherwise, the piece should travel along the longest path (the bottom right panel).
(b) Simplified board with 20 sites considered in the present paper. Only the longest path in (a)
is allowed, which makes the game board effectively one dimensional.
}
\label{fig:YutPan}
\end{figure}

Pieces move along the sites marked on the board counterclockwise, starting
from the bottom right corner.  Allowed
sites are drawn as circles, and there are three forks, two upper corners and
center [see Fig.~\ref{fig:YutPan}(a)].  There are four possible ways to return to
the starting point to make a round trip; the shortest path,
two middle distance paths, and the longest path as shown
in Fig.~\ref{fig:YutPan}(a).  Rules of the game are almost
common across the whole country with a small variation for different
regions.  Each team, one by one in turn, rolls the sticks and moves pieces
on the board.  The team, whose all pieces make round trips around the board
returning to the starting point, is the winner.  If other team's pieces are
at the site of the stick-rolling team's piece, other team's pieces are
caught and the team is asked to roll sticks one more time.  We call this
situation as {\em catching} in the present paper.  Yut and Mo have less
chance to occur in comparison with Do, Gae, and Geol (see Fig.~\ref{fig:Pn})
and when the team has either Yut or Mo, the team is given another
chance to throw sticks.  Consequently, one team often can
throw sticks several times in one turn.  Suppose that a team gets Yut,
and thus the team throws sticks once more to get Mo, which gives the team
Geol in next rolling.  If the last Geol catches the opponent piece(s), the
team can throw once more.  Consequently, there can occur a variety of
different situations even in a single turn, and it is possible in principle
that one team can win the game in its first turn although present authors
have not seen this in their entire lives.  If the same team's pieces meet at the
same site on the board, they can pile up (we call this {\em piling}).  Piled
up pieces can move together afterwards so that piling reduces the number of
turns that is needed to win.  On the other hand,
piling can be risky, since if piled pieces are caught by an opponent team,
all pieces should start the game again from the starting position.
Accordingly, catching is a more aggressive strategy in Yut-Nori, but piling
is also a good strategy in some situation in which the chance
of being caught is slim.

The original two-dimensional game board [see Fig.~\ref{fig:YutPan}(a)] is too
complicated to analyze. Only for the sake of simplicity, we consider only the
longest path in the game board, which allows us to analyze the one-dimensional
board displayed in Fig.~\ref{fig:YutPan}(b).  The advantage of using the
one-dimensional board lies on the fact that we do not need to choose which path
to take among the four possibilities in Fig.~\ref{fig:YutPan}(a). Within the
limitation, we believe
that our game theoretic approach described below can give us some clue to
answer the question which strategy between piling and catching yields bigger winning rate.

The remaining part of the present paper is organized as follows: In
Sec.~\ref{sec:1D}, we first consider the case in which one player with one
piece throws Yut sticks. We perform a simple statistical analysis of our
simplified one-dimensional Yut-Nori  to compute the average step size
and the distribution of number of turns to finish the game. Our main result for
the winning strategy is reported in Sec.~\ref{sec:winning}, and the avalanche-size
distribution in the infinitely long one-dimensional board is discussed in
Sec.~\ref{sec:avalanche}, which is followed by summary in
Sec.~\ref{sec:summary}.

\section{Movement of a piece in one-dimensional Yut-Nori}
\label{sec:1D}

We first consider the simplest case that one player plays the game alone
with only one piece in one-dimensional Yut-Pan [see Fig.~\ref{fig:YutPan}(b)].
We can ask how long distance the piece advances in one turn on average.
Let us denote this expectation value of the distance as $\Phi$. Since
the player can throw Yut sticks once more if she has Yut or Mo, $\Phi$ satisfies
the following self-consistent equation
\begin{equation}
\Phi = P(1) + 2 P(2) + 3 P(3) + (4+\Phi)P(4) + (5+\Phi)P(5),
\end{equation}
where $P(m)$ is the probability for the step size $m$, e.g., $m=3$ for Geol
and $m=5$ for Mo, respectively, and we get
\begin{equation}
\label{eq:Phi}
\Phi = \frac{4p + 5(1-p)^4}{1-p^4 - (1-p)^4}.
\end{equation}
When $p=0.6$ (see Fig.~\ref{fig:Pn}), Eq.~(\ref{eq:Phi}) yields
$\Phi = 2.992424$, which we confirm through numerical simulation:
From $10^6$ runs, the average value is found to be $\Phi = 2.992(2)$.

\begin{figure}
\includegraphics[width=0.5\textwidth]{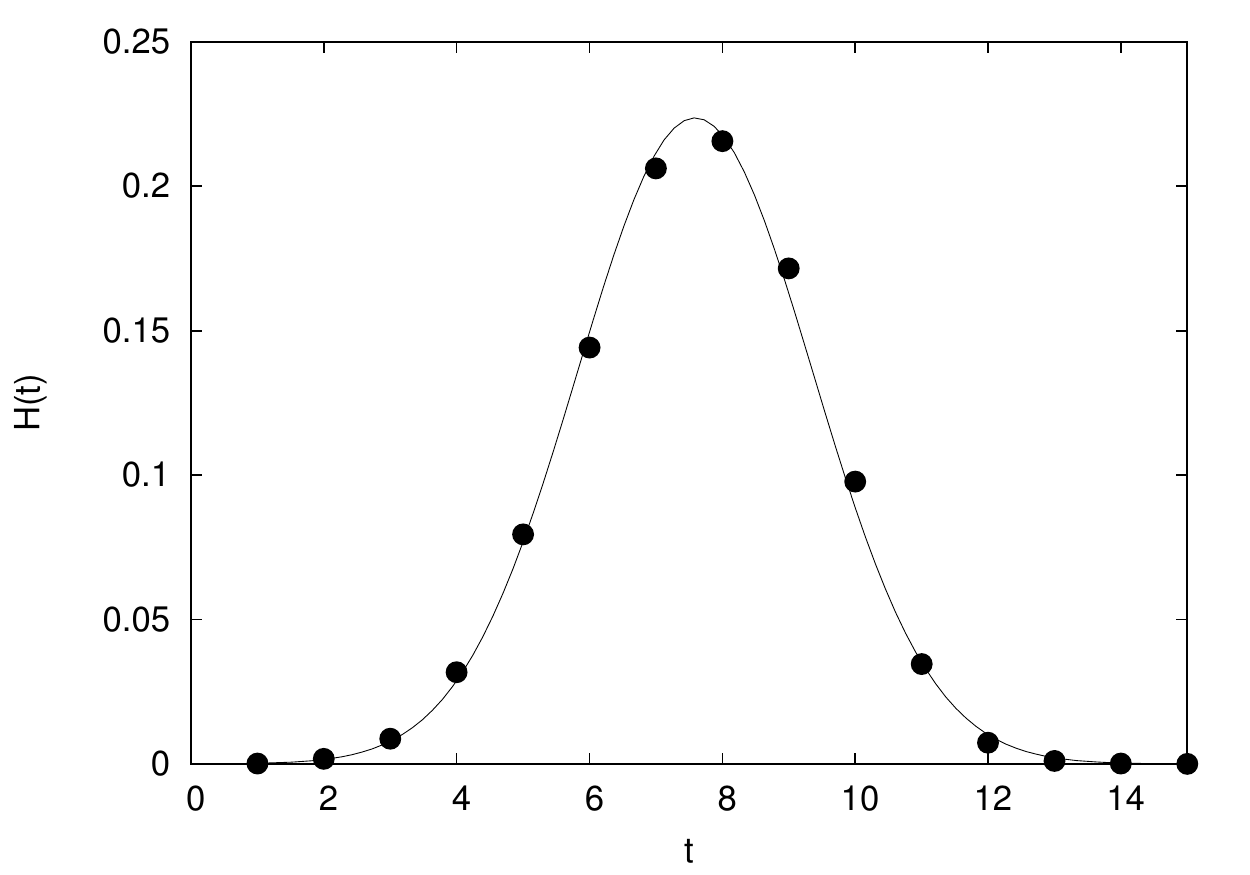}
\caption{The distribution function $H(t)$ versus $t$ (the number of turns to finish
the game) calculated from $10^6$ independent runs (filled circles).
The full-drawn curve is the outcome from the least-square fit to the Gaussian form
with the mean 7.59 and the standard deviation 1.77.}
\label{fig:gauss}
\end{figure}

During numerical experiment repeated $10^6$ times, we also
measure how many turns needs to be spent to finish the game and call it $t$.
The distribution function $H(t)$ is very well approximated by
the Gaussian form as shown in Fig.~\ref{fig:gauss}. Among the $10^6$ runs,
we find $t=1$ occurs only about 100 times.
Likewise,
in playing the original Yut-Nori with the game board in Fig.~\ref{fig:YutPan}(a),
it is almost impossible to see the starting team win the game without ever
giving its turn to other teams.

\section{Winning strategy}
\label{sec:winning}

Although more than two teams can play Yut-Nori, two-team play appears
to be most popular. In every round, each team throws sticks in turn,
and the player who actually throws sticks is chosen sequentially:
Suppose that two teams $A$ and $B$ have players $A_1, A_2,
A_3$ and $B_1, B_2, B_3$, respectively. In the first round $A_1$ and $B_1$
throw sticks, and in the second round $A_2$ and $B_2$ play, and so on.
In our setting, the number of players in each team is not important if
all players in the team use the same strategy.
To further simplify the game, we assume that two teams use only two pieces each
(instead of four as in the original Yut-Nori) and play the game
in the one-dimensional board in Fig.~\ref{fig:YutPan}(b). Even in this simple
setting a variety of different situations can arise.
Team's strategy is crucial to win the game since there are different ways to
move pieces on the board even in the same situation.
A team can be aggressive and tries to catch the opponent's pieces whenever
possible. Otherwise, it can pile up their pieces to speed up the movement of pieces afterwards.

Our simplified game proceeds as follows:
A team throws Yut sticks. Depending on the outcome (see Fig.~\ref{fig:yut}),
the number $m$ of steps the team's piece can advance is decided.
For given $m$, the team can decide what to do for the current
configuration of the game board. If the piece of the opponent can be caught for
given $m$ we denote this situation as '$C$', while we denote '$P$'
if the team can pile up their two pieces together. If neither catching nor
piling is possible, we denote this as '$N$'. Since each team has two
pieces, the team's strategy must decide what to do for all possible cases
$(X_1, X_2)$ with $X_1, X_2 = C, P, N$,
where $X_{1(2)}$ is the situation for the first (second) piece.
For example, $(X_1,X_2) = (C,P)$ means that the team's first piece can catch
the opponent's piece and 
the team's second piece can pile up with its first piece.
We denote the decision made by a given strategy as $S(X_1,X_2)$. If
the strategy $S$ decides to move the 1st piece, $S(X_1,X_2) = (X_1,0)$,
while $S(X_1,X_2) = (0,X_2)$ if the 2nd piece is decided to be moved.
For example $S(C,P) = (C,0)$ means that the strategy $S$
chooses to catch by using the first piece, while $S(C,P) = (0,P)$ means
that the 2nd piece will be moved to pile up with the 1st piece.
Among all $9 (=3 \times 3)$ possible combinations for $(X_1,X_2)$, it is clear
that $(P, P)$ is impossible since only one piece that is behind the other piece
can pile up. Among remaining 8 cases,
$(P,N)$, $(C,N)$, $(N,P)$, and $(N,C)$ are easy to handle: We naturally
assume that teams always favor $P$ and $C$ over $N$, giving us
$S(P,N) = (P,0)$, $S(C,N) = (C,0)$, $S(N,P) = (0,P)$, $S(N,C) = (0,C)$,
irrespective of given strategy. For $(C,C)$, it is always better to catch
the opponent's piece that is located close to the destination and thus
$S(C,C) = (C,0)$ or $(0,C)$ depending on the locations of opponent's pieces.
From the above consideration, the only cases to be taken carefully are
$(P,C)$, $(C,P)$, and $(N,N)$. Among these three, $(N,N)$ is not related
with the choice between catching and piling, and thus we fix the strategy
outcome for this: We move the piece which has bigger chance to be caught
by the opponent team. Suppose that the distance to the opponent team's piece behind
is $d$. We put the highest priority for the piece if $d=2$ or $3$, and the second
highest priority for $d=1$, and then for $d=4$ (see Fig.~\ref{fig:Pn}).
If $d \geq 5$, we move the piece that is behind of the other piece.

In the present work, we aim to find which
strategy between catching and piling gives bigger chance to win, and thus
consider two different strategies denoted by $S_C$ and $S_P$. The former
prefers catching to piling, and the latter piling to catching. Accordingly,
$S_C(C,P) = (C,0)$ and $S_C(P,C) = (0,C)$, whereas $S_P(C,P) = (0,P)$ and
$S_P(P,C) = (P,0)$.  We also need to consider the effect of who starts
first, which makes us consider all four different combinations
of competition: $[S_1, S_2] = [S_C, S_C], [S_C, S_P], [S_P, S_C], [S_P, S_P]$,
where $S_{1(2)}$ is the strategy chosen by the first (the second) team.
We calculate the wining rate of the first team from $10^9$ independent runs,
and report the result in Table~\ref{tab:win}.

We find that if the 2nd team's strategy is $S_C$, the 1st team is better off
if $S_C$ is used (the top left corner of Table~\ref{tab:win}). If the 2nd
team is using $S_P$ instead, the 1st team again has a higher winning rate for $S_C$.
We thus
conclude that $S_C$ always gives better chance to win regardless of the
opponent's strategy. Other seemingly interesting finding one can draw from
Table~\ref{tab:win} is that always the first team has bigger chance to win.
In other words, being the first team to throw Yut sticks is more important
than the strategy choice, within the limitation of simplification made
in the present work.

For more detailed understanding of the dominance of $S_C$, let us compare
advantages of two strategies, $S_C$ and $S_P$, in terms of dominance. We define
dominance of one team, say $A$, against the other team, say $B$, as
\begin{equation}
    D=x^A_1+x^A_2-x^B_1-x^B_2,
\end{equation}
where $x_1$ and $x_2$ denote positions of two pieces of each team. If the
number of steps given to the team $A$ is $m$ and the team $A$'s situation is
$(P,C)$, the corresponding configuration must be such that $x^A_1+m=x^A_2$ and
$x^A_2+m=x^B_1$ (suppose that the 1st piece of $B$ can be caught).
In case when the team $A$'s strategy
is $S_C$, the piece at $x^A_2$ catches the piece at $x^B_{1}$ so that the
expected increment of $D$ is written as $\Delta D_C = m+x^B_1+\Phi$. Here
$\Phi$ is added due to the fact that the team $A$ is given one more
chance to throw sticks. If
the team $A$ uses the strategy $S_P$, then the piece at $x^A_1$ will be piled
on the top of the piece at $x^A_2$. Piling itself does not contribute to the
increment of $D$ at this turn. Instead, the team $A$ will benefit from the
piling at later turns unless they are caught by the piece of team $B$ behind
them. Thus, we get $\Delta D_P=m+c\Phi$, where $c$ is the expected number of
free movement of the piled piece. The condition for the dominance of $S_C$,
i.e. $\Delta D_C>\Delta D_P$, reads $x^B_1>(c-1)\Phi$. This inequality holds
only for sufficiently large $x^A_2$ because larger $x^A_2$ leads to both larger
$x^B_1$, which is $x^A_2+m$, and smaller $c$. Precisely, the expectation of $m$
is $\Phi$, and one gets $c\Phi=20-x^A_2$ if the team $A$ is not interrupted by
the team $B$ after piling, resulting in the condition as $x^A_2>10-\Phi\approx
7$ for $p\approx 0.6$. Despite of the oversimplification, our argument gives a
hint for understanding the results.

To conclude the present section, in the
present numerical experiment of strategy competition,
an aggressive team who starts the game first has a higher chance to win.

\begin{table}
\caption{Winning rates for the 1st team for different
pairs of competing strategies. $S_C$ prefers
catching the opponent's piece to piling up pieces, while $S_P$'s
preference is opposite (it favors piling over catching). If the 1st
team uses $S_C$, the 2nd $S_P$, the 1st team wins the game
at probability 0.51314(2). It is found that you'd better be
the first team and that $S_C$ always gives higher winning rates
than $S_P$.
We repeat $10^9$ runs to get the average winning rates.
The numbers in parentheses are the errors in the last digits.
}
\begin{tabular}{c|cc}\hline
\backslashbox{2nd}{1st}  &$S_C$ & $S_P$\\\hline
$S_C$  & 	$0.51314(2)$   &  $0.50883(2)$ \\\hline
$S_P$ & 	$0.51629(2)$   &  $0.51274(2)$ \\\hline
\end{tabular}
\label{tab:win}
\end{table}


\section{Avalanche-size distribution}
\label{sec:avalanche}

Suppose that there are many pieces on the board. If the density
of opponent team's pieces is very high, your piece can easily catch
those pieces, giving you one more chance to throw Yut sticks and thus
to catch more. Accordingly, one can expect that if the density
of opponent team's pieces is high, the density will soon
be reduced. On the other hand, if the density is low, it
will gradually increase because most of pieces will survive. The above
reasoning suggests a possibility that the board configuration
can be self-organized to some state, which might be critical (or not).

We construct a simple model of the one-dimensional game with infinite number
of teams and infinite number of sites.  We restrict each team to have only
one piece, and the piece can move only one (at probability $p$)
or two (at probability $1-p$) steps.
If the piece is decided to move two steps at probability $1-p$,
or if it catches other piece,
the team has one more chance to move. This simplified model can be
considered as Yut-Nori with one Yut stick, with only Do and Mo.
Of course, differently from the original Yut-Nori, Mo with no belly
sided stick gives a team two steps, instead of five steps, in this
simplified version of the game.

\begin{figure}
\includegraphics[width=0.45\textwidth]{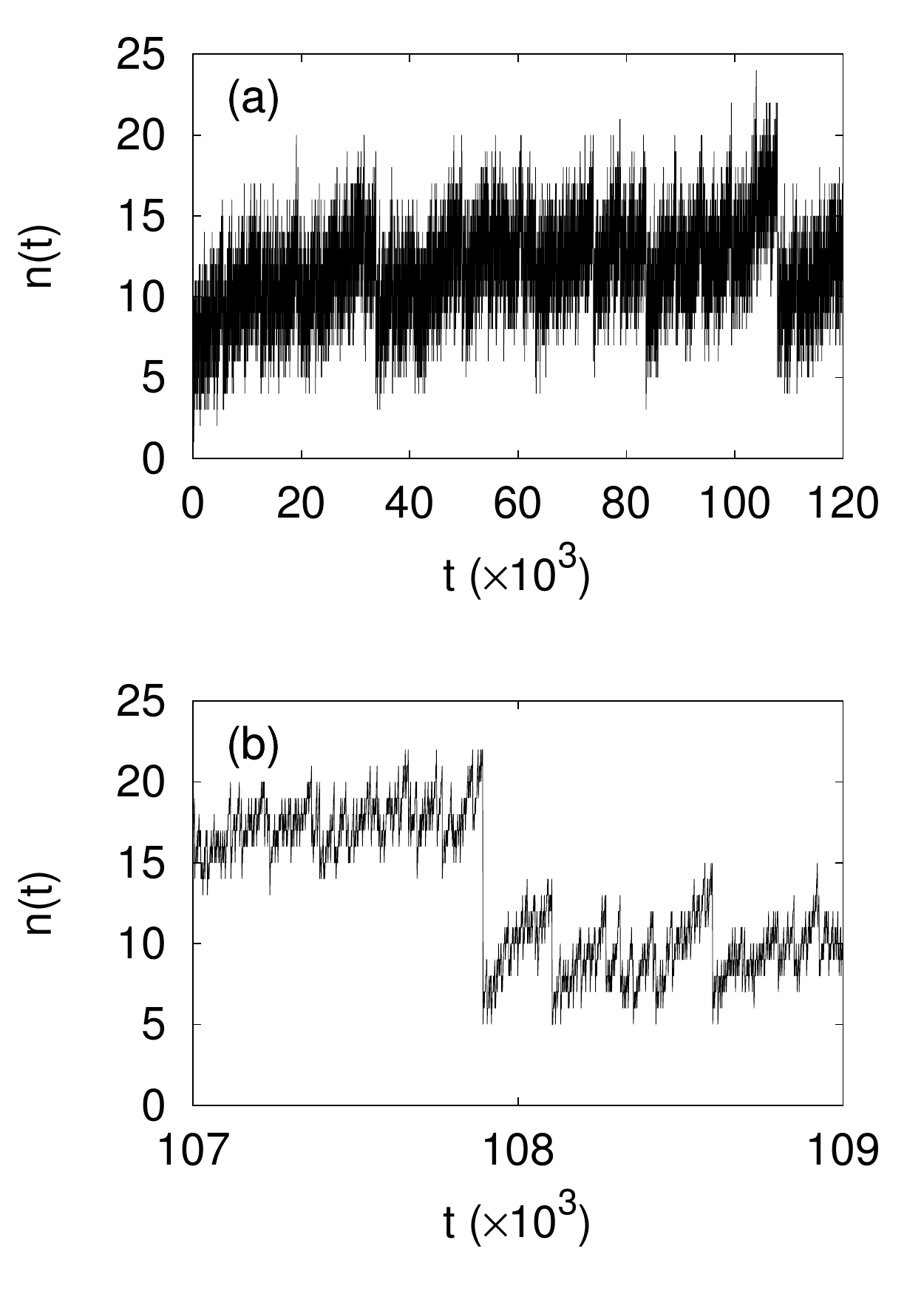}
\caption{(a) The number $n(t)$ of pieces on the board at time $t$ for
$p=0.8$. (b) Expansion of (a) showing abrupt drops of $n(t)$. The
avalanche size $s$ is defined as the drop size whose distribution
is displayed in Fig.~\ref{fig:aval}.
}
\label{fig:density}
\end{figure}

\begin{figure}
\includegraphics[width=0.5\textwidth]{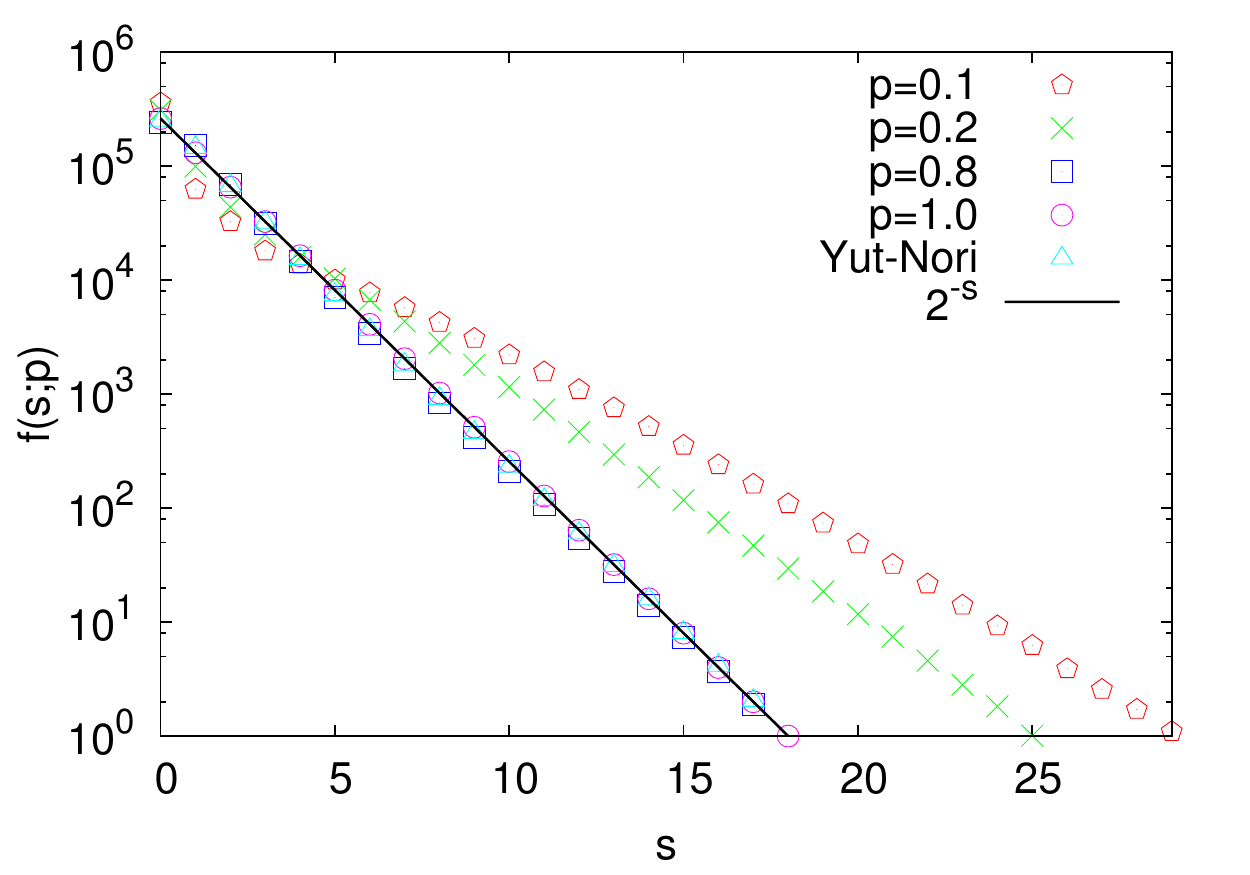}
\caption{Distribution function $f(s;p)$ of the avalanche size $s$
for the simplified Yut-Nori on infinitely long one-dimensional board.
We assume that there is only one Yut stick, and if the stick shows
belly side up (at probability $p$) the piece moves one step. Otherwise,
for the back side, the piece is moved two steps and the team is
given one more chance to play, like Mo in the original game.
We also assume that each team is given one piece and the number of teams
is not limited, which means that every piece on the board is put by
a different team.
For any value of the probability $p$ for the belly side,
the avalanche-size distribution clearly  follows an exponential form:
$f(s; p) \sim e^{-s/\xi_p}$ with $\xi_p$ decreasing with $p$.
At $p=1$, $f(s;p=1) \sim 2^{-s}$, as explained in the text.
We also include the result (triangles) for the four stick game (like the original Yut-Nori),
which lies almost on top of curves for $p=1$ and $p=0.8$.
}
\label{fig:aval}
\end{figure}

The number of pieces on the board varies in time.  It first increases with
fluctuation [see Fig.~\ref{fig:density}(a)], and it sometimes decreases drastically [see
Fig.~\ref{fig:density}(b)].
Avalanche size $s$ is defined as
the number of caught pieces removed from the board in one turn, and
we display the avalanche-size distribution (see Ref.~\cite{CC} for the
a general discussion) in Fig.~\ref{fig:aval}.
Irrespective of $p$, the avalanche-size distribution
is shown to follow an exponential form $f(s; p) \sim e^{-s/\xi_p}$
with $\xi_p$ decreasing with $p$.

For $p=1$, the process becomes trivially deterministic, which allows us to get
$f(s;p=1)$ analytically.  The first player starts and put her piece at the
first site. We represent the configuration of the board as "$\cdots 000000001$"
(the first site is the rightmost one), which we denote as binary digit 1.
Likewise, after the second turn we get the board "$\cdots 00000010$" $=10$, and
$11$ for the third turn, $100$ for the fourth turn, $101$ for the fifth turn,
all in the binary format.  In this binary notation, the board configuration
after the $m$th turn is written as a binary number which is simply $m$ in
decimal format.  Consequently, in order to occupy all sites up to $n$th site,
one needs $2^{n}-1$ turns since the next player at the $2^n$-th turn will have
her piece at the $(n+1)$-th site removing all $n$ pieces on its way. In other
words, the avalanche of the size $s=n$ occurs at time $t=2^n$.
Avalanches that occur
in the time interval $t \in [2^{n-1}+1, 2^n-1]$ are exactly identical to the
avalanches in $t \in [0, 2^{n-1}-1]$.  It is interesting to see that the
above process is similar to the game of Hanoi towers, in which one is asked to
move all disks to other pole within the constraint that a smaller disk must be
put on a larger disk.
For $t < 2^n$,
the largest avalanche was of the size $n-1$ and it occurred once.  The second
largest avalanche size was $n-2$ and it occurred twice, and the avalanche of
the size $n-3$ occurred four times, and so on.
Consequently, the distribution of the avalanche sizes $s$ is written as
\begin{equation}
\label{eq:Hanoi}
f(s, p = 1) \sim 2^{-s}.
\end{equation}
In Fig.~\ref{fig:aval}, it is seen that for $p \gtrsim 0.8$, the avalanche-size
distribution almost overlaps with $2^{-s}$ in Eq.~(\ref{eq:Hanoi}).
We also present the avalanche-size distribution obtained for four Yut-stick
game like the original Yut-Nori in Fig.~\ref{fig:aval}, which also
exhibits exponential form and well described by $2^{-s}$ in Eq.~(\ref{eq:Hanoi}).
In four Yut-stick game, Yut and Mo occur at probability about 0.15 in total, which
presumably explains the agreement with the one stick game at $p \approx 0.85$,
as shown in Fig.~\ref{fig:aval}, since Yut and Mo give players a chance
to rethrow sticks.

It has been well known that one-dimension Abelian sandpile models do
not show self-organized criticality~\cite{1DAbelian} whereas
the avalanche-size distribution of the non-Abelian model follows
a power law form~\cite{1DnonAbelian}.
Our Yut-Nori model is close to the one-dimensional non-Abelian model~\cite{1DnonAbelian}.
However, the avalanche-size distribution of our model does not follow
the power law because  dissipation of pieces can occur at any place in the board.
In the presence of severe dissipation of pieces as in our model,
the average avalanche size was shown not to diverge and the distribution has
a cutoff~\cite{CC}, which explains the absence of the self-organized criticality in our model.

\section{Summary}
\label{sec:summary}

One-dimensional version of the Korean game of  Yut-Nori has been
investigated. It has been shown that the probability $p$ for one Yut stick
to show the flat belly side up must be close to 0.6. When Yut-Nori
is played, there can be two competing ways to decide how to move pieces
on the game board Yut-Pan; A team can either prefer piling, or catching.
By performing numerical simulations of huge number of simplified
virtual Yut-Nori games,
we have shown that the more aggressive strategy that favors catching yields
a higher winning rate. It has been also found that the team who starts first
has a higher chance to win. The avalanche-size distribution has been investigated
and has been shown not to exhibit self-organized criticality. The exponential
form of the distribution has been understood from the consideration of
the limiting case that Yut stick always shows belly side up.

\begin{acknowledgments}
B.J.K. was
supported by the National Research Foundation of Korea (NRF) grant funded by the
Korea government (MEST) (No. 2011-0015731). H.J. acknowledges
financial support by Aalto University postdoctoral program.
\end{acknowledgments}

\bibliographystyle{apsrev4-1}

\end{document}